# Gallium nanoparticles grow where light is

K. F. MacDonald*, W. S. Brocklesby[†], V. I. Emel'yanov[‡], V. A. Fedotov*, S. Pochon*, K. J. Ross*, G. Stevens* and N. I. Zheludev*.

*Department of Physics and Astronomy, University of Southampton, Highfield, Southampton, Hampshire, SO17 1BJ, U. K.*
*e-mail: kfm@phys.soton.ac.uk*

[†] *Optoelectronics Research Centre, University of Southampton, Highfield, Southampton, Hampshire, SO17 1BJ, U. K.*

[‡]*Department of Physics, Moscow State University, Moscow 119899, Russia*

**The study of metallic nanoparticles has a long tradition in linear and nonlinear optics[1], with current emphasis on the ultrafast dynamics, size, shape and collective effects in their optical response[2] [3] [4] [5] [6]. Nanoparticles also represent the ultimate confined geometry: high surface-to-volume ratios lead to local field enhancements and a range of dramatic modifications of the material's properties and phase diagram[7] [8] [9]. Confined gallium has become a subject of special interest as the light-induced structural phase transition recently observed in gallium films[10] [11] has allowed for the demonstration of all-optical switching devices that operate at low laser power[12]. Spontaneous self-assembly has been the main approach to the preparation of nanoparticles (for a review see 13). Here we report that light can dramatically influence the nanoparticle self-assembly process: illumination of a substrate exposed to a beam of gallium atoms results in the formation of nanoparticles with a relatively narrow size distribution. Very low light intensities, below the threshold for thermally-induced evaporation, exert considerable control over nanoparticle formation through non-thermal atomic desorption induced by electronic excitation.**

We studied nanoparticle formation at the end of an optical fibre exposed to a beam of gallium atoms in vacuum (see fig. 1). This arrangement allowed us to accurately control the deposition conditions and simultaneously to probe the optical properties of the developing structures. Our experiments were conducted in a vacuum chamber evacuated to ~$10^{-6}$ mbar during the deposition process. We used single-mode silica optical fibres, with core and cladding diameters of 9 and 125 μm respectively. One end of the fibre was freshly cleaved and attached to a cold-finger in the chamber that could be cooled by liquid nitrogen flow to 100 K. It was exposed to a beam of gallium atoms from a K-cell to form a semi-reflective layer on the cleaved face of the fibre. In most of our experiments Gallium was deposited at a rate of ~0.3 nm/min (measured with a quartz balance). The other end of the fibre (outside the vacuum chamber) was connected to a diode laser operating at 1.5 μm, producing 1 μs pulses with peak powers of 17 mW at a rate of 1 kHz (an average power of 17 μW). After deposition fibres were removed from the vacuum chamber and examined using optical and Atomic Force Microscopes (AFM).

We first deposited gallium on a fibre tip at room temperature for thirty minutes in the absence of any laser light and found that a glossy feature-less mirror-like layer was formed (AFM scans, with resolution down to a few nanometers, showed that the surface was flat). In our second experiment the fibre was cooled to 100 K. In this case, thirty minutes of deposition (again without light present) produced a uniform, rough layer of gallium. An AFM scan of the fibre tip revealed that the gallium was in lumps of all shapes and sizes from a few nanometers to several hundred nanometers. Next we performed deposition on a fibre cooled to 100 K with the laser switched on. Thirty minutes of deposition now led to the formation of a layer with two distinct parts (see photograph fig. 2a, and AFM image fig. 2b). Outside the fibre's core the structure was identical to that produced in the previous experiment with no laser light present, but at the core a highly reflective area was formed. Further AFM scans revealed that this area comprised a layer of gallium nanoparticles of relatively uniform size, typically ~80 nm in diameter (fig. 2c). We found that if the fibre tip was more than ~1 mm from the supporting surface of the cold finger, it was not effectively cooled and the bright area at the core was not formed. To determine whether the light-assisted nanoparticle formation took place during deposition or during warming to room temperature we performed two experiments (both with thirty minutes of deposition on fibres at 100 K): one with the laser on during deposition and off while warming the fibre; and another with light absent during deposition but present while the fibre was warmed. Only in the first of these experiments, with light during deposition, was a well-defined reflective area of nanoparticles formed at the fibre's core. Finally, we replaced the laser with a 0.8 mW continuous wave diode laser operating at 1.31 $\mu$m and once more deposited gallium for thirty minutes on a fibre cooled to 100 K. This laser had some effect on the gallium, however it produced only a diffuse reflective area at the fibre's core rather than the well-defined highly reflective region produced by the pulsed laser. In summary, we found that the presence of laser pulses during deposition and efficient cooling of the fibre tip are essential for the creation of nanoparticles at the core.

The difference between areas illuminated during deposition and those not illuminated becomes even more obvious when gallium is deposited at a very low rate for several hours. In this case the large blobs of gallium separated by tens of microns form in areas not exposed to the lasers but at the fibre's core an isolated cluster of well defined nanoparticles develops (see fig. 3).

Our experiments indicate that the following sequence of events takes place during formation of gallium nanoparticles in an illuminated area. Gallium, deposited at a low rate (~0.3 nm/min) initially forms seeding clusters on the surface of the silica substrate through surface diffusion and nucleation. Since gallium wets silica in vacuum rather well[14] the initial shape of the nucleated particles formed at low deposition rates has a high area-to-height aspect ratio[15]. The clusters grow preferentially across the surface, increasing their aspect ratio and space fill factor. A reasonable model for calculating their optical properties can be based on cylinders of gallium with radii bigger than their height, or highly truncated spheres. Due to depolarization effects[16] and collective dipole-dipole interactions between particles and substrate[17] the absorption cross-section of the nanoparticle[18] depends on the dielectric properties of both the substrate and the particle bulk material, filling factor and particle size, but most importantly on the shape of the nanoparticle, i.e. aspect ratio. Although we do not know precisely the crystalline phase of the particles, their dielectric parameters must lie between those of semi-metallic α-gallium, which is the stable bulk phase under normal conditions[19], and those of crystalline phases such as Ga-II, Ga-III and β-Ga which have free-electron-like optical properties[20][21]. We therefore calculated, using the theory developed in references 16-18, the absorption cross-section of α- and free-electron gallium nanoparticles with different aspect ratios. In α-gallium, atoms form covalently bound dimers which were assumed in our calculations to be aligned perpendicular to the substrate. This is the preferred orientation for gallium dimers at a hard wall[22], although even without this assumption the results of the calculations remain largely the same. Data on the dispersion of the dielectric coefficients in α-gallium were taken from Kofman *et al*[23], and for free-electron gallium from Hunderi and Ryberg[24]. The calculations show that in both cases the absorption cross-section at 1.55μm increases rapidly with aspect ratio, and thus with the size of the particles (fig. 4).

Consequently, laser excitation in the infrared part of the spectrum will show better coupling with bigger particles - they will be more strongly excited (see fig. 4a). This leads to a higher rate of desorption of gallium atoms from bigger particles. As noted by Vollmer and Trager[25], atoms located at the edges or perimeters of nanoparticles have particularly low binding energies and will evaporate preferentially. This size-selective evaporation will terminate the growth process in bigger particles. Therefore, when light is present in the fibre, it slows down the formation of bigger particles and prevents the formation of particles above a certain size, but has little effect on the formation of smaller particles. Consequently the particle size distribution remains narrow. This model is consistent with the layers' optical transmission measurements which were taken in-situ immediately after deposition. The transmission, at 1.55μm, of the mirror exposed to light during deposition was about 18 times higher than that of an unexposed mirror. This is related to the fact that large gallium nanoparticles, which are more strongly absorbing, are not created when light is present during deposition, but can form in the absence of light.

Important features of our experiment are the low intensity and relatively long duration of the optical pulses used. We modeled the temperature increase $\Delta T$ resulting from light absorption in our experiments by numerically solving the thermal conductivity equation for a disk-like nanoparticle using the Green's function technique[26] and found the maximum increase to be ~80 K. This relatively small increase cannot affect the evaporation rate. Indeed, in gallium nanoparticles the activation energy for atomic desorption is 2.5 eV [ref. 27]. This is more than 100 times higher then the maximum value of $k_b T$, making thermal desorption, whose rate is proportional to $exp(-E_g/k_b T)$, a forbidden process. We believe that the increased rate of desorption in our experiments is the consequence of a non-thermal process related to electronic excitation of the particles. In literature this mechanism is called DIET: Desorption Induced by Electronic Transitions, and has been studied in potassium, sodium and silver[28]. This non-thermal mechanism should be especially important for α-gallium because its crystalline structure contains covalently bound gallium dimers, $Ga_2$ molecules.

Importantly, gallium's bonding-antibonding absorption line encompasses the 1.3 and 1.55 μm wavelengths used in our experiments[29]. Our results suggest that when optical stimulation excites a dimer from the bonding to the antibonding state the activation energy of atomic desorbtion from the nanoparticle is dramatically reduced, thus allowing a non-thermal particle size control mechanism to operate at very low light intensity levels.

When the deposition process is completed and the fibre is brought to room temperature, the clusters melt and form the slightly truncated spheres seen in the AFM images. Comparison of the transmission characteristics of the mirrors immediately after deposition and shortly after bringing the mirror to room temperature, i.e. after melting of gallium islands is consistent with this model. Indeed, the transmission increases by about 60% for mirrors exposed to light during deposition, and by factor of 25 for mirrors that were not. Clearly, as absorption is higher for higher aspect ratios, when the gallium islands with high aspect ratios are converted into truncated spheres with low aspect ratios on melting, there is a greater change in absorption for mirrors which initially had particles with high aspect ratios, i.e. mirrors which had not been exposed to light during deposition. An additional increase in the particles' sphericity may be expected when the mirror is brought into contact with air. Such a change in the shape is due to the higher wetting angles (due to oxidation) between gallium and silica under normal atmospheric conditions, as compared to vacuum conditions[17].

It is worth comparing our optical excitation energy levels with those used for *post-growth* narrowing of the size distribution of silver and gold nanoparticles[30]. In size-selective evaporation of silver clusters the energy densities used were 100 and 150 mJ/cm$^2$, corresponding to peak intensities on the silver target of 14 and 21 MW/cm$^2$. In our experiments with light-controlled deposition, some effect on particle formation is seen with continuous laser excitation at intensities of just 1.2 kW/cm$^2$, about four orders of magnitude lower than the levels used in post-growth narrowing. With pulsed laser excitation we observe a formidable light-assisted nanoparticle formation effect at levels of 28 kW/cm$^2$, still about 700 times lower than the post-growth narrowing levels.

In conclusion, we report that the formation of metallic nanoparticles can be induced at very low intensity levels by light-induced desorption of atoms during the process of deposition from an atomic beam. We expect that by changing the excitation wavelength or the atomic beam intensity, the size and spatial distribution of gallium nanoparticles can be controlled.

**Figure 1**
Sketch of the apparatus for light-assisted self-assembly of gallium nanoparticles on the end of an optical fibre.

**Figure 2**
(a) Optical microscope image, in reflection, of a fibre end after 30 minutes of gallium deposition at ~0.3 nm/min. The bright spot in the center (~9 μm in diameter) corresponds to the fibre's core where light assists the self-assembly of nanoparticles; (b) AFM image of an area encompassing the fibre's core. The core is seen as a "crater"; (c) AFM image of nanoparticles inside the core area (illuminated during deposition). The peak-to-valley range is 61 nm; (d) Area occupied by particles as a function of their size for image c. The most common particle size is 80 nm with a standard deviation of 22 nm (the finite size of the probe tip is not taken into account); (e) AFM image of an area outside the core (not illuminated during deposition) showing irregular formations of gallium. The peak-to-valley range is 348 nm; (f) Area occupied by particles as a function of their size for image e, showing the broad range of sizes present.

**Figure 3**
(a) Optical microscope image, in reflection, of a fibre end after 360 minutes of gallium deposition at a rate of ~0.1 nm/min. The total mass of gallium on this fibre is four times the mass deposited on the fibre shown in fig.'s 2a-f. The spot in the center corresponds to the fibre's core where a cluster of gallium nanoparticles is formed; (b) AFM image of an area encompassing the fibre's core, showing an isolated cluster of nanoparticles at the core (dashed line).

**Figure 4**
(a) Calculated optical absorption cross-section $\sigma$ per unit volume $V$ at 1.55 μm for a gallium particle on a silica substrate as a function of aspect ratio $D/h$ (see inset). Data are shown for three different filling factors $q$ - the fraction of the total surface covered by gallium. Solid and dashed lines correspond to α-gallium and free-electron gallium respectively; (b) Calculated spectral dependencies of the optical absorption cross-section $\sigma$ of α-gallium particles of identical height but different aspect ratio on a silica substrate, assuming a cylindrical form. Dipole-dipole interactions with the substrate and with neighboring particles were accounted for in our calculations. A rapid increase in absorption cross-section is seen with increasing aspect ratio, and therefore with increasing particle size.

Figure 1

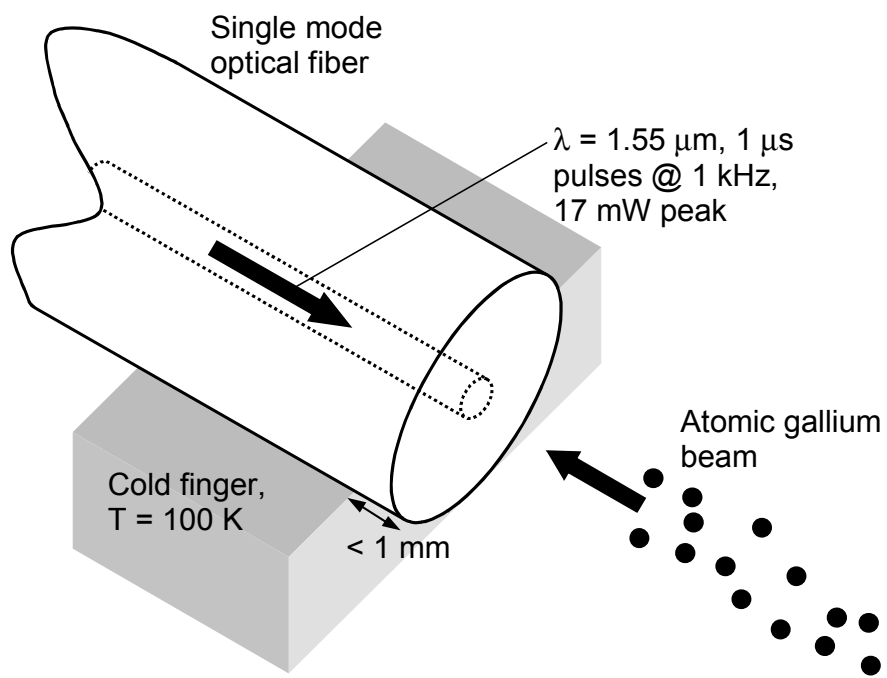

Single mode
optical fiber

λ = 1.55 μm, 1 μs
pulses @ 1 kHz,
17 mW peak

Atomic gallium
beam

Cold finger,
T = 100 K

< 1 mm

Figure 2

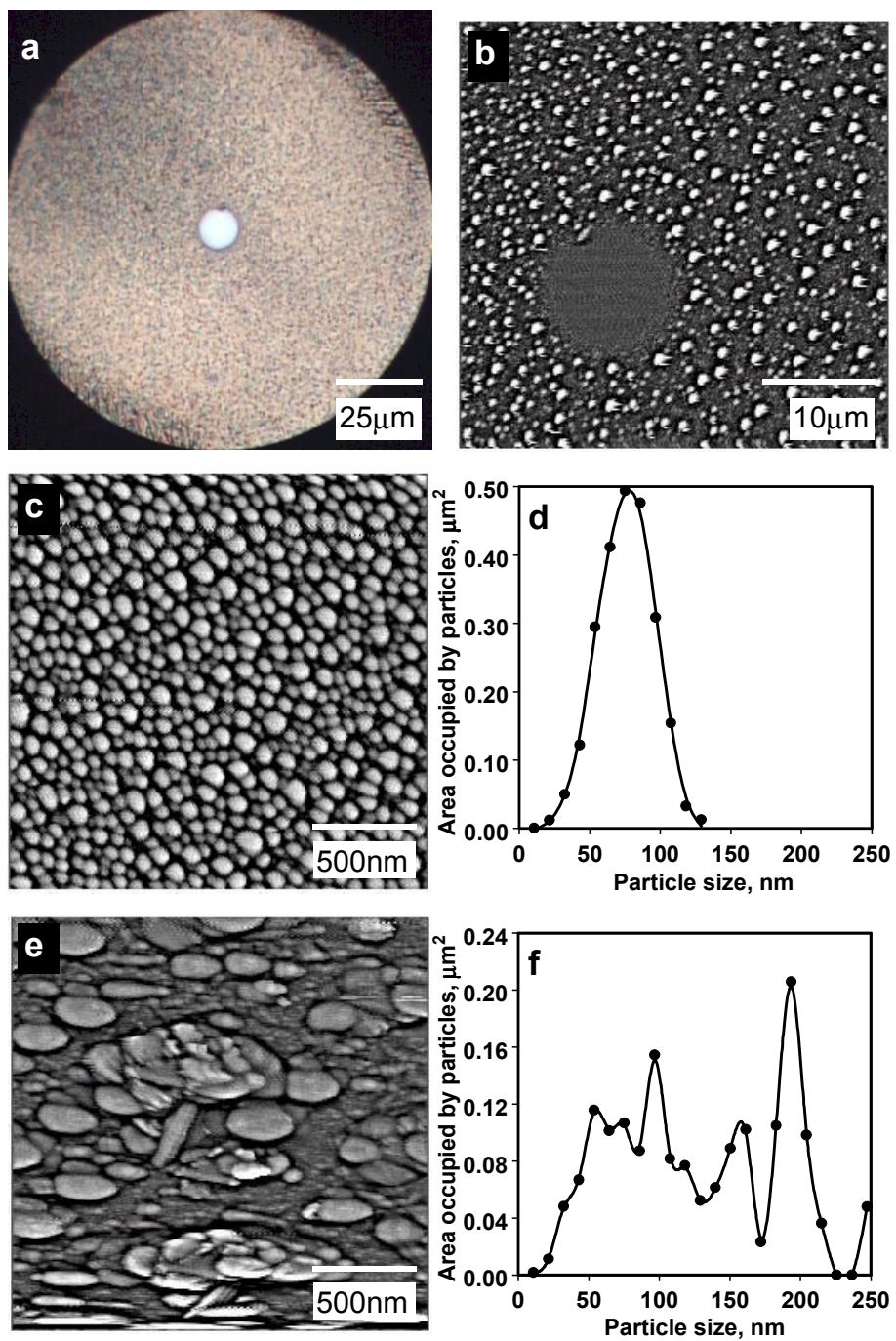

Figure 3

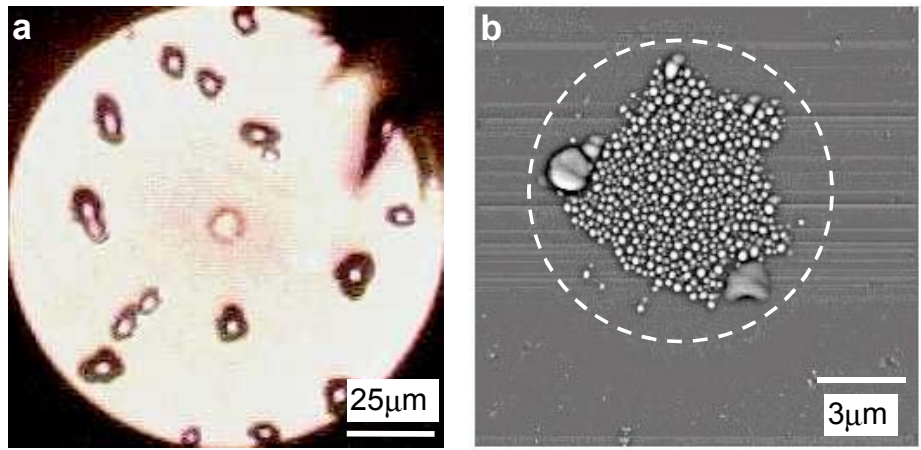

Figure 4

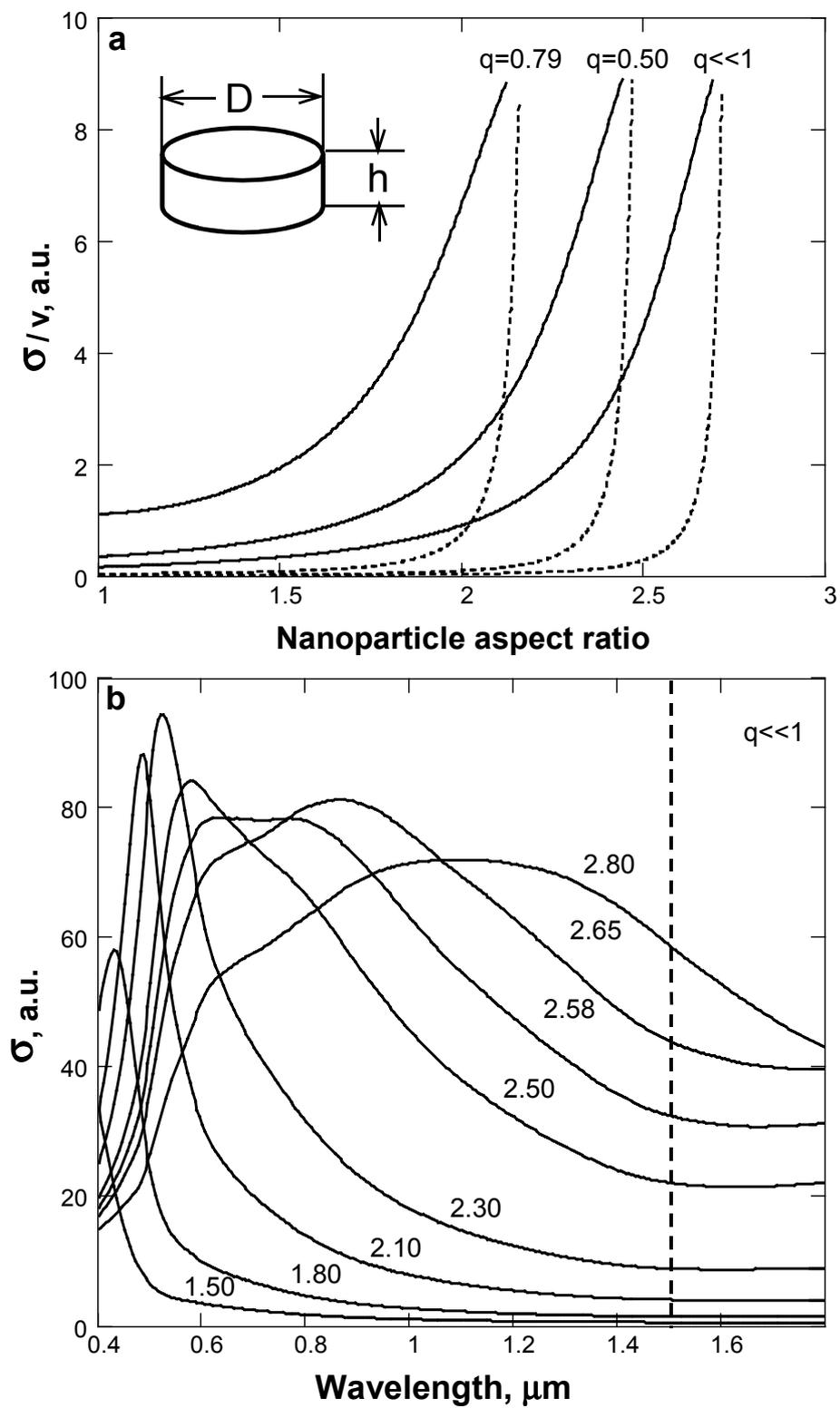